# SENSITIVITY OF FINITE ELEMENT MODELS TO RELATIONSHIP BETWEEN $T_2$ RELAXATION AND MODULUS IN ARTICULAR CARTILAGE

[1] Alexander A. Donabedian
[1,2]† Deva D. Chan

[1] Weldon School of Biomedical Engineering, Purdue University, West Lafayette, IN 47907
[2] School of Mechanical Engineering, Purdue University, West Lafayette, IN 47907

†Corresponding Author: Deva Chan (chand@purdue.edu)

Key Terms: Articular Cartilage, Finite Element Analysis Sensitivity, $T_2$ Relaxation, Dynamic Modulus

**ABSTRACT**

Correlating articular cartilage material properties to quantitative magnetic resonance imaging biomarkers is a powerful approach to biofidelic finite element models. However, subject-specific relationships between imaging biomarkers such as $T_2$ and material properties like dynamic modulus are uncertain. To evaluate the sensitivity of finite element models to this uncertainty, we shifted the slope and intercept of a linear $T_2$-dynamic modulus relationship used to define cartilage properties. Modulus shifts led to notable percent changes in the top 1% of calculated stress and strain while modulating slope had a negligible impact, together supporting the use of physiologically relevant moduli ranges in subject-specific models.

# INTRODUCTION

Articular cartilage lines the ends of bones in synovial joints enabling near frictionless contact while aiding in the transfer of loads to the subchondral bone during daily movement. Osteoarthritis is a degenerative joint disease that targets the tissue's content and structure (Pritzker, *et al.* 2006, Roberts, *et al.* 1986, Saarakkala, *et al.* 2010) which impedes healthy function. Due to the poor healing qualities of articular cartilage (Gomoll and Minas 2014), identifying OA at the early stages and assessing drug interventions can provide clinicians with valuable insight towards prognosis. While noninvasive imaging, such as magnetic resonance imaging (MRI) can distinguish between healthy and patients with OA (Dunn, *et al.* 2004, Zhao, *et al.* 2022), biomechanical models can simulate *in vivo* conditions and could be predictive of future damage. Finite element (FE) models can compute the short-term elastic (Carey, *et al.* 2014, Mootanah, *et al.* 2014, Orsi, *et al.* 2016) and long-term biphasic response (Mattei, *et al.* 2014, Meng, *et al.* 2014) of cartilage to address its complex nature. The material depth-dependency (Buckley, *et al.* 2008) and variability between subjects (Peters, *et al.* 2018) as well as content changes associated with OA (Roberts, *et al.* 1986, Saarakkala, *et al.* 2010) underscores the importance of implementing subject-specific material properties to improve the model's biofidelity. While some studies implement specimen-specific collagen microstructure (Pierce, *et al.* 2010), fixed charge density (Rasanen, *et al.* 2016) or tissue compositional volume fractions (Linka, *et al.* 2019), most models are a function of properties based on literature and specific material properties remains an uncommon practice.

While the direct measurement of tissue material properties is ideal to inform subject-specific FE models, noninvasive estimation is advantageous for potential use in the clinic. Although, material estimation relies on the relationship between noninvasive outputs to tissue mechanics which is not trivial. Inverse approaches can be used to estimate modulus from displacement-encoded MRI (Butz, *et al.* 2011) and magnetic resonance elastography can estimate the shear modulus through propagating shear waves (Muthupillai, et al. 1995). Despite these advancements, these applications are constrained to research environments because customized equipment and pulse sequences are necessary for these applications which are currently impractical to implement in the clinic.

On the other hand, quantitative MRI is a well-established and clinically feasible technique that correlates to tissue content and material properties. Transverse relaxation times ($T_2$) correlate with articular cartilage water content (Chou, et al. 2009) and collagen architecture (Nieminen, et al. 2001). In collagenase-digested bovine cartilage, $T_2$ increased while stiffness decreased compared to control (Nieminen, et al. 2000), reinforcing the idea of a negative relationship between $T_2$ times and cartilage strength. The equilibrium and dynamic modulus were also strongly negatively correlated to bulk $T_2$ of human articular cartilage collected from various regions of the patella (Lammentausta, et al. 2006). In the weight-bearing regions of articular cartilage, this linear relationship between $T_2$ and both moduli hold, albeit with greater variability (Nissi, et al. 2007). Lampen, et al., leveraged this linear relationship between $T_2$ and modulus to develop subject-specific $T_2$-refined FE modeling approach, wherein the dynamic modulus of articular cartilage scaled, by element, with corresponding $T_2$ (Lampen, et al. 2023). However, the high variability of the experimental data from multiple studies that relate cartilage $T_2$ and moduli (Kurkijarvi, et al. 2004, Lammentausta, et al. 2006, Nissi, et al. 2007) suggests that the $T_2$-modulus relationship may itself be subject-specific, calling into question whether the same $T_2$-modulus relationship could be used to define FE models of different individuals. Thus, it is necessary to evaluate the degree to which image-based FE models such as those introduced by Lampen, et al., are sensitive to the relationship defined between a quantitative MRI metric and tissue material properties.

In this study, we assessed the sensitivity of $T_2$-informed FE models to the linear relationship used to assign dynamic modulus based on local $T_2$. Our primary objective was to measure changes in key model outputs resulting from changes to the input $T_2$-modulus relationship in our previously established $T_2$-informed FE approach (Lampen, *et al.* 2023). To accomplish this, we also compared two different approaches to assigning modulus to each element based on the overlapping $T_2$ maps.

## METHODS

### *Imaging Data Pre-Processing*

We used data from the Osteoarthritis Initiative (OAI), a 10-year observational study aimed to study knee OA (Peterfy, *et al.* 2008). The OAI dataset includes baseline MRI with incremental 12-month follow-up exams from 4,796 participants. Four subjects with available high-quality hexahedral meshes were selected from a study by Rodriguez, *et al.* (Rodriguez-Vila, *et al.* 2017). Multi-echo-time scans for these subjects were available to estimate transverse relaxation time ($T_2$) with in-plane spatial resolution of 0.3125 mm × 0.3125 mm and slice thickness of 3.48 mm. $T_2$ maps were calculated with 3D Slicer (Slicer 5.0.2, Kitware, Clifton Park, N. Y.)(Fedorov, *et al.* 2012) and smoothed using an edge preserving gradient anisotropic diffusion filter in SimpleITK (Yaniv, *et al.* 2018) with five iterations, time step=0.125, and conductance = 3 to reduce large $T_2$ discrepancies between adjacent voxels.

### *$T_2$-Informed Finite Element Model*

All simulations were preprocessed using The Geometry and Image-Based Bioengineering add-ON (GIBBON) toolbox (Moerman 2018) in MATLAB (R2024a, The Mathworks, Natick, MA) prior to FE analysis in FEBio Studio version 2.5.0 (Maas, *et al.* 2012). Each mesh consisted of hexahedral elements making up the articular cartilage and menisci morphology segmented from dual echo and steady state (DESS) scans available through OAI (Rodriguez-Vila, *et al.* 2017). The compressible neo-Hookean material model implemented previously (Lampen, *et al.* 2023) was used to model a hyperelastic and isotropic response. Cartilage was assumed nearly incompressible with Poisson's ratio of ν = 0.45 and dynamic modulus ($E_D$) scaled according to the $T_2$-$E_D$ relationship, detailed below. The meniscus was assumed to be homogeneous with E = 20 MPa and ν = 0.3 (Mononen, *et al.* 2015).

Boundary conditions were defined based on *in situ* loading of articular cartilage and menisci. The femoral cartilage-femur interface was assumed to be effectively rigid, with nodes at the bone interface of femoral cartilage modeled as fixed to a rigid body. The center of mass of the rigid body was defined as the midpoint of line connecting the femoral epicondyles. The nodes at the bone interface of the tibial cartilage were fixed using a zero-displacement boundary condition. Linear springs, with a spring constant equal to that of the menisci (20 MPa), were attached at the anterior and posterior horns of the menisci to model meniscal root attachment to the tibia. Frictionless contact was assumed between the articular cartilage and menisci surfaces. An average motion and loading condition from the stance phase of a walking gait cycle (Bergmann, *et al.* 2014) were applied through the center of mass. All models incorporated an initial ramping phase of 0.1 s, followed by a loaded stance phase of 0.9 s. During the ramping phase, the loading conditions were incrementally applied until the endpoint matched the starting conditions of the stance phase. Positioning of the femoral cartilage of each subject was manually adjusted to ensure continuous contact with the tibial cartilage during the full loading cycle.

For each subject, the $E_D$ of each element of articular cartilage was scaled according to a corresponding $T_2$ map. The $T_2$ map was first transformed to the same Cartesian coordinate system as the model and manually registered using a rigid transformation (Figure 1A). Because the resolution mismatch between mesh elements and $T_2$ map voxels results in multiple voxels overlapping a single element (Figure 1B), we implemented two different approaches to assign a $T_2$ to each element: a nearest-neighbor or volume-weighted approach. The nearest-neighbor approach assigned a $T_2$ value to each element based on the nearest voxel by centroid-to-centroid distance (Lampen, *et al.* 2023) using the TransformPhysicalPointToIndex function in SimpleITK (Figure 1B).

The volume-weighted approach assigned a $T_2$ value to each element ($T_{2_e}$) that was calculated based on overlapping volume fraction ($f_V$) (Equation 1).

$$T_{2_e} = \sum T_2 \times f_V \qquad (Equation\ 1)$$

To convert from $T_2$ to $E_D$, a negative linear relationship derived from experimental data (Nissi, *et al.* 2007) was fit to the range of $T_2$ values within the OAI dataset (Equation 2), to assign each element a modulus, as previously described (Lampen, *et al.* 2023).

$$E_D = \left(\frac{-3.5}{3} \times 10^5\right) T_{2_e} + 9.75 \times 10^6 \qquad (Equation\ 2)$$

Since all subjects used in this study had minimal signs of OA at the time when the $T_2$ maps were taken, a 15 to 75 ms $T_2$ range was chosen to align with the $E_D$ range expected from articular cartilage with minimal OA (approximately 1-8 MPa) (Waldstein, *et al.* 2016). Any voxel with a higher or lower $T_2$ time was set to upper or lower bound, respectively.

To assess the sensitivity of the FE model to changes in the input $T_2$-$E_D$ relationship, a baseline model for each subject was created with the default $T_2$-$E_D$ relationship (Equation 2) using the volume-weighted approach. Then, additional models were created with two types of changes to the $T_2$-$E_D$ relationship: modulus shift and altered slope (Figure 1C). For "modulus shift," the $E_D$ range was shifted up or down incrementally by 10% to cover the full range of dynamic moduli from previous experiments (Nissi, *et al.* 2007) while maintaining the same slope. For "altered slope," the slope of the $T_2$-$E_D$ relationship was altered incrementally by 10%, holding the midpoint of the default $T_2$-$E_D$ relationship fixed. In the latter conditions, a positive change in slope moves the model towards a homogeneous $E_D$. The model parameters for each subject-specific model otherwise remained constant.

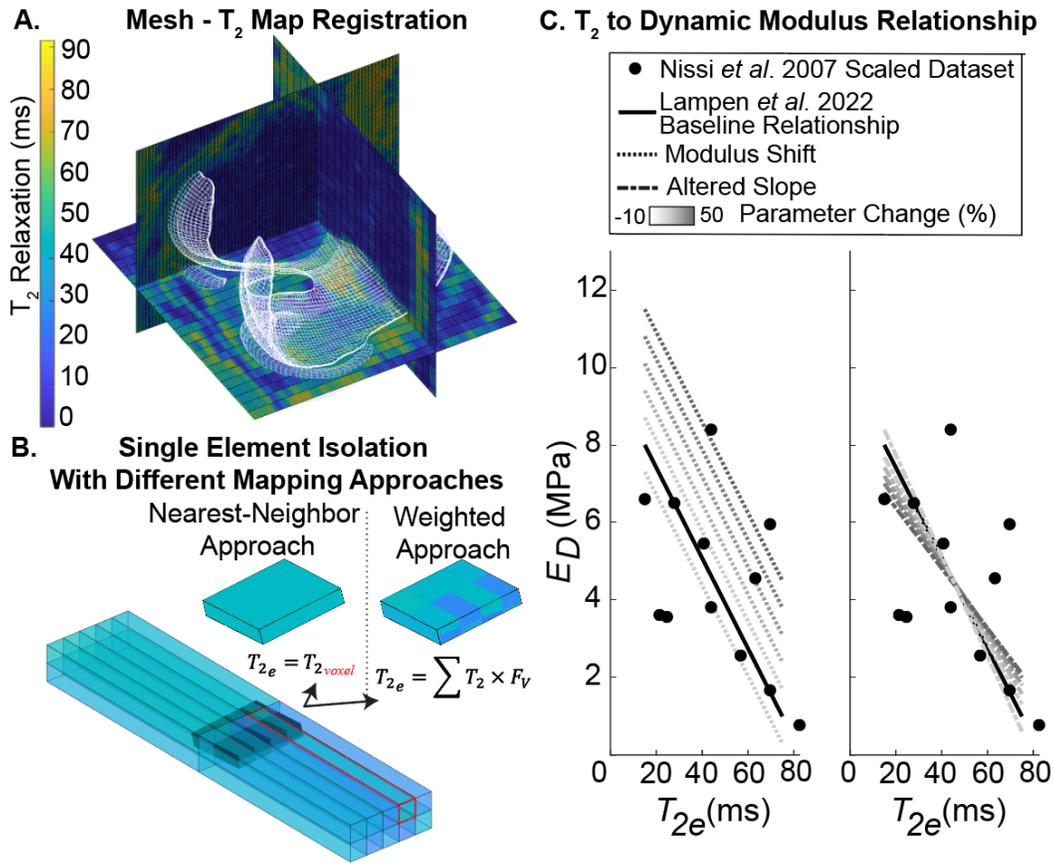

**Figure 1. Magnetic resonance imaging $T_2$ to element dynamic modulus ($E_D$) pipeline.** A) The subject-specific mesh and corresponding $T_2$ map were registered. B) $T_2$ values were assigned to an element ($T_{2e}$) following one of two approaches. The nearest-neighbor approach assigned $T_{2e}$ based on the voxel closest to the element centroid, as indicated by the voxel outlined in red. The weighted approach summed $T_2$ scaled by its volume fraction, $F_V$, over all overlapping voxels. C) The $T_2$-$E_D$ relationship (Nissi, *et al.* 2007) was confined to the range of $T_2$ values found in the OAI dataset (solid black line). For the sensitivity study, the intercept or slope were adjusted by -10 to 50%. For a "modulus shift," the $E_D$ range was adjusted while maintaining the same slope to $T_2$. For an "altered slope," the $T_2$-$E_D$ slope was adjusted about the mean of the midpoint of the baseline relationship.

## *Intrasubject Comparisons*

The $T_2$ maps assigned using nearest-neighbor and volume-weighted approaches were qualitatively and quantitatively compared. Scatter and Bland-Altman plots were used to determine the correlation and the limits of agreement between the two approaches. The roughness and root mean square (RMS) of the $T_2$ mappings from both approaches were calculated, and the outcomes were compared using paired t-test. Statistical significance was defined at $p < 0.05$.

The sensitivity of the FE model to the input $T_2$-$E_D$ relationship was evaluated using model outputs at different phases of stance. The first principal stress (most tensile), third principal stress (most compressive), and maximum shear stress resulting from modulus shift or altered slope were evaluated and compared to the default model. The analysis focusses on the top 1% of stress and strain values from all simulations as osteoarthritic lesions can be linked back to infrequent yet high mechanical stress (Seedholm, *et al.* 1979). The number of elements under stress greater than the baseline third principal stress and max shear stress were compared among the modulus shift and altered slope extremes using histograms. The percent change in both principal and shear outputs was calculated with respect to the default $T_2$-$E_D$ relationship at heel strike, midstance, and heel off and compared to the degree of parameter relationship change. The overall sensitivity to either modulus shifts or altered slopes was determined by the behavior of the output percent change compared to the degree of parametric adjustment.

**RESULTS**

The FE models for 4 subjects from the OAI dataset showed a sensitivity to the $T_2$-$E_D$ relationship intercept, but not the slope. Additionally, the assigned mesh $T_2$ values depended on the mapping method.

*Voxel to Element $T_2$ Assignment*

Qualitative and quantitative comparisons of the assigned $T_2$ values from the volume-weighted approach differed in time and surface texture compared to that from the nearest neighbor approach (Figure 2A). The nearest neighbor approach assigned $T_2$ values for the 21936 to 27912 elements per subject almost instantaneously, but the volume-weighted approach required ~10 hours per subject.

The weighted approach resulted in a smoother $T_2$ mapping compared to the nearest neighbor approach (Figures 2B-C). The $T_2$ roughness and RMS calculated in the femoral and tibial cartilage separately, as well as both regions combined, were significantly lower using the volume-weighted approach compared to the nearest neighbor approach (Table 1). Both approaches produced similar $T_2$ assignments per element indicated by the $R^2$ value of 0.724 (Figure 2D) and fell within mutual limits of agreement (Figure 2E).

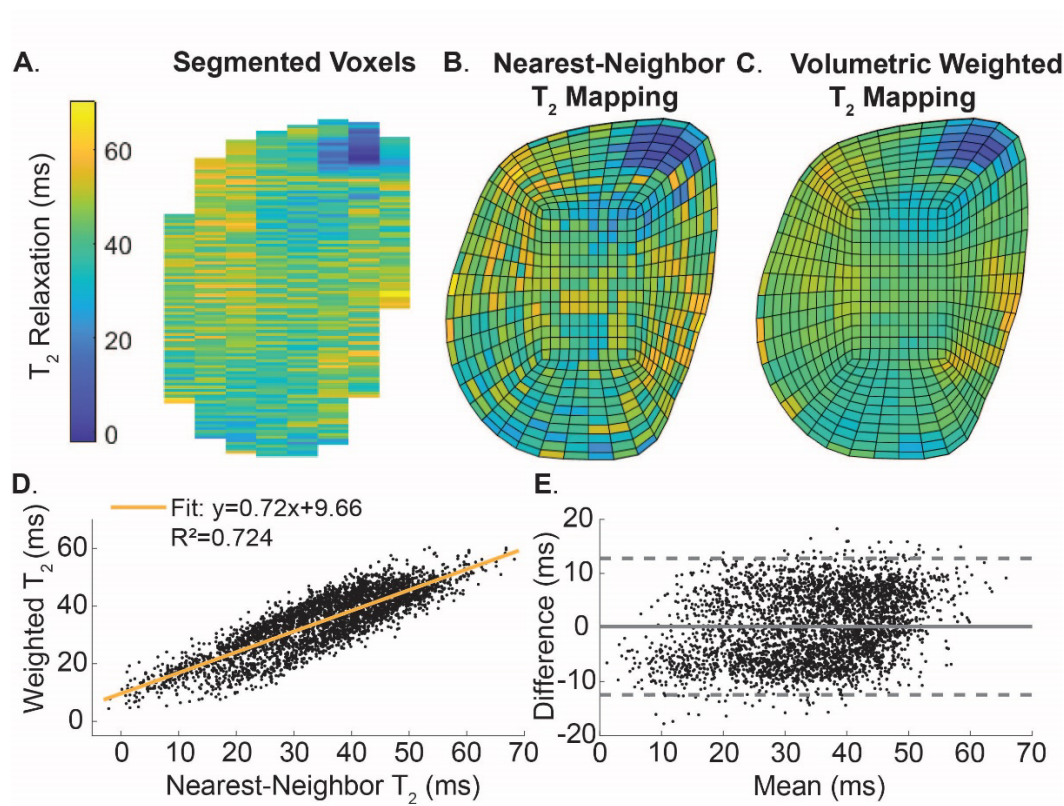

**Figure 2. Representative comparisons between nearest-neighbor and volume-weighted approaches to assign $T_2$ values to elements.** A) All voxels used to inform the elements making up the lateral tibial cartilage of Subject 9932809 were identified. $T_2$ was assigned to each element using the B) nearest neighbor or C) weighted approach. D) The $T_2$ value assigned to each element from both approaches correlate well. E) Bland-Altman plots showed that both approaches produced $T_2$ assignments within limits of agreement.

**Table 1. T₂ texture resulting from T₂ assignment approaches.** The $T_2$ roughness and root mean square (RMS) were calculated in all, femoral-only, and tibial-only articular cartilage elements, following $T_2$ assignment by the nearest-neighbor or weighted approach. These metrics across four subjects (mean ± standard deviation (SD)) were calculated and compared by paired t-tests, showing that the weighted approach consistently resulted in smoother $T_2$ distributions (*, p<0.05).

|  | Articular Cartilage $T_2$ [ms] | | | Femoral Cartilage $T_2$ [ms] | | | Tibial Cartilage $T_2$ [ms] | | |
|---|---|---|---|---|---|---|---|---|---|
|  | Nearest Neighbor | Weighted | p | Nearest Neighbor | Weighted | p | Nearest Neighbor | Weighted | p |
| **Roughness** | 11.7 ± 1.2 | 9.4 ± 1.1 | *0.042* | 10.4 ± 1.1 | 8.1 ± 0.2 | *0.023* | 10.8 ± 1.1 | 8.6 ± 1.6 | *0.014* |
| **RMS** | 14.5 ± 1.4 | 11.7 ± 1.4 | *0.034* | 13.1 ± 1.4 | 10.5 ± 0.5 | *0.030* | 13.3 ± 1.3 | 10.5 ± 1.7 | *0.012* |

## *Sensitivity of the $T_2$-$E_D$ Relationship*

Third principal stress and maximum shear stress were compared for sensitivity studies involving maximum and minimum modulus shift (Figure 3) and altered slope (Figure 4) of the $T_2$-$E_D$ relationship as well as intermediate parameter perturbations (Supplemental Figures S1-S4). The top 1% values of the first principal, third principal, and maximum shear stresses (Figure 5) and strains (Figure 6) at heel strike, midstance, and heel off were compared for all 4 subjects (Supplemental Tables S1 and S2).

Altering slope did not substantially impact the computed outputs (Figure 3). A 50% altered slope revealed 23 and 39 fewer elements exceeding the baseline top 1% of the third principal and shear stress respectively in the representative subject 9961728 (Figure 3). Additionally, the altered slope led to at most a 0.4, -3.5%, and -3.5% mean change in the first principal stress, third principal stress, and max shear stress top 1% values respectively between all subjects (Figure 5; Supplemental Table 1) as well as a -2.6%, -2.1%, and a -2.3% mean change in the first principal strain, third principal strain, and max shear strain top 1% values respectively between all subjects (Figure 6; Supplemental Table 2).

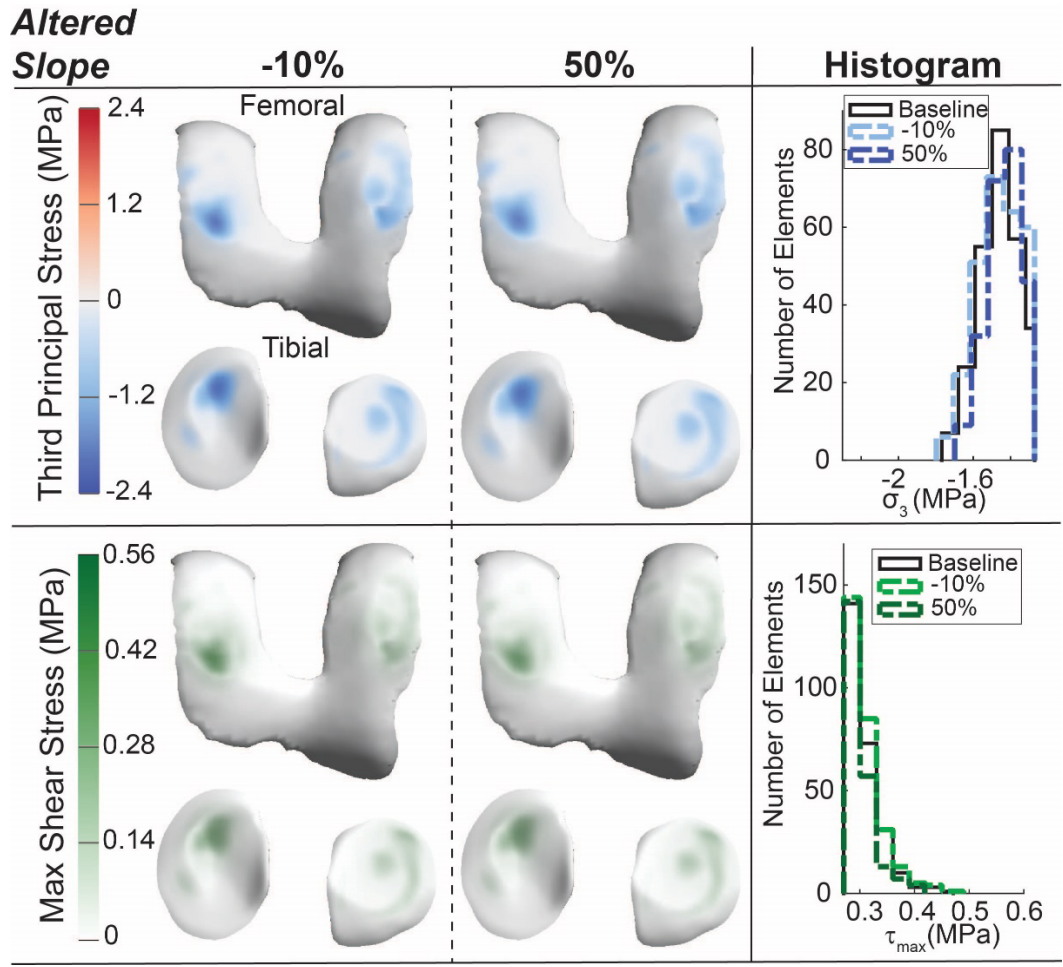

**Figure 3. Mechanical response as a result from altered slope.** Third principal stress and shear stress at midstance were calculated for models with the maximum decreased (-10%) and increased (+50%) slope for a representative subject (9961728). A histogram of the elements with stresses greater than the top 1% values from the baseline simulation show minimal changes in response to altered slope.

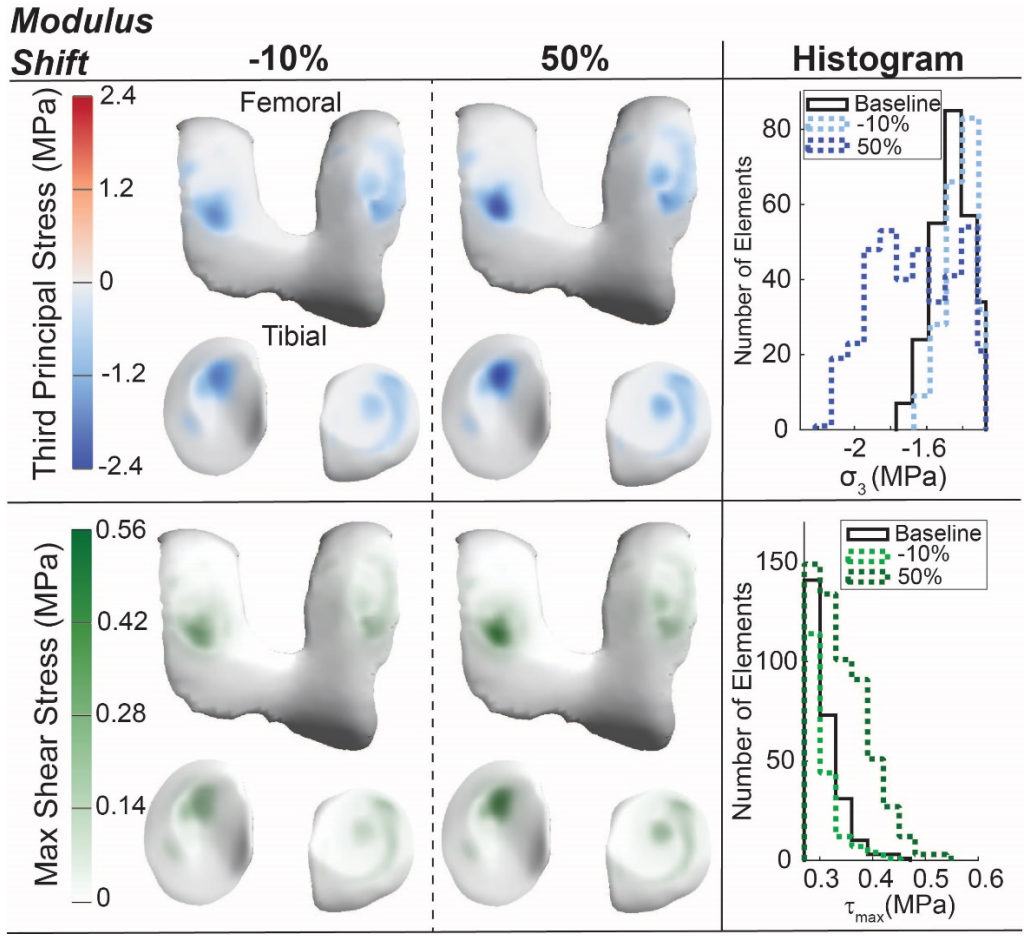

**Figure 4. Mechanical response as a result from the modulus shift.** Third principal stress and shear stress at midstance were calculated for models with the maximum decreased (-10%) and increased (+50%) modulus shift for a representative subject (9961728). A histogram of elements with stresses greater than the top 1% values from the baseline simulation show larger changes in response to modulus shift.

Shifting modulus up increased the number of elements under higher stress (Figure 4), increased the top 1% stress values (Figure 5), and decreased the top 1% strain values (Figure 6). A 50% modulus shift resulted in 120 and 310 more elements exceeding the baseline top 1% third principal stress max shear stress values respectively from subject 9961728 (Figure 4). The maximum modulus shift also led up to a 32.1%, 24.7%, and a 30.8% mean change in the first principal stress, third principal stress, and max shear stress top 1% values between all subjects respectively (Figure 5; Supplemental Table S1) as well as a -28.5%, -25.6%, and a -26.2% mean change in the first principal strain, third principal strain, and max shear strain top 1% values between all subjects respectively (Figure 6; Supplemental Table S2).

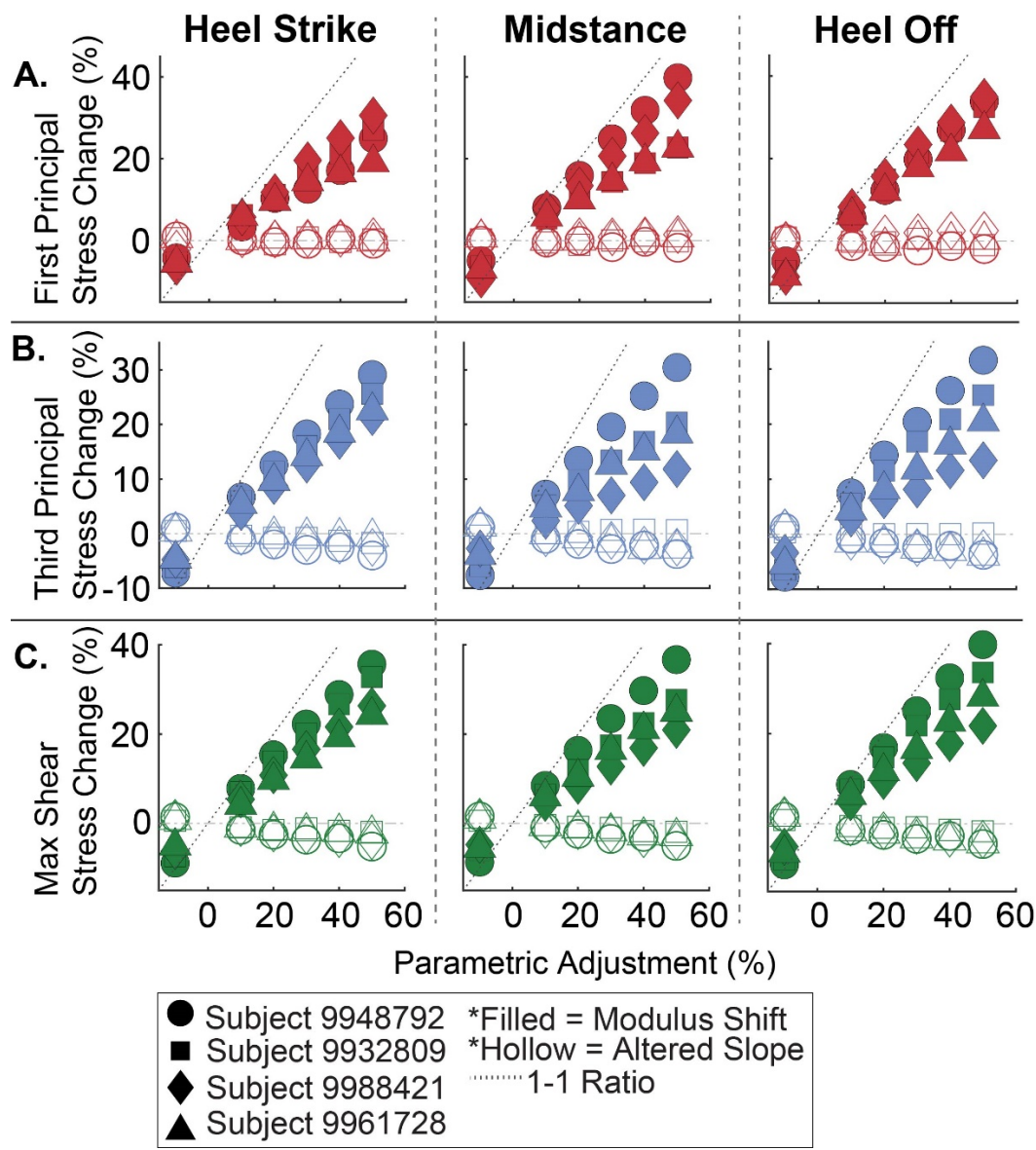

**Figure 5. Percent change in the key stress outputs after $T_2$-$E_D$ perturbation.** The percent change in the top 1% stress metrics at heel strike, midstance, and heel off was computed between the baseline simulation and simulations implementing 10% incremental changes to the $T_2$-$E_D$ relationship parameters. Solid shapes represent the modulus shift whereas hollow shapes as a result from the changes to the slope relationship.

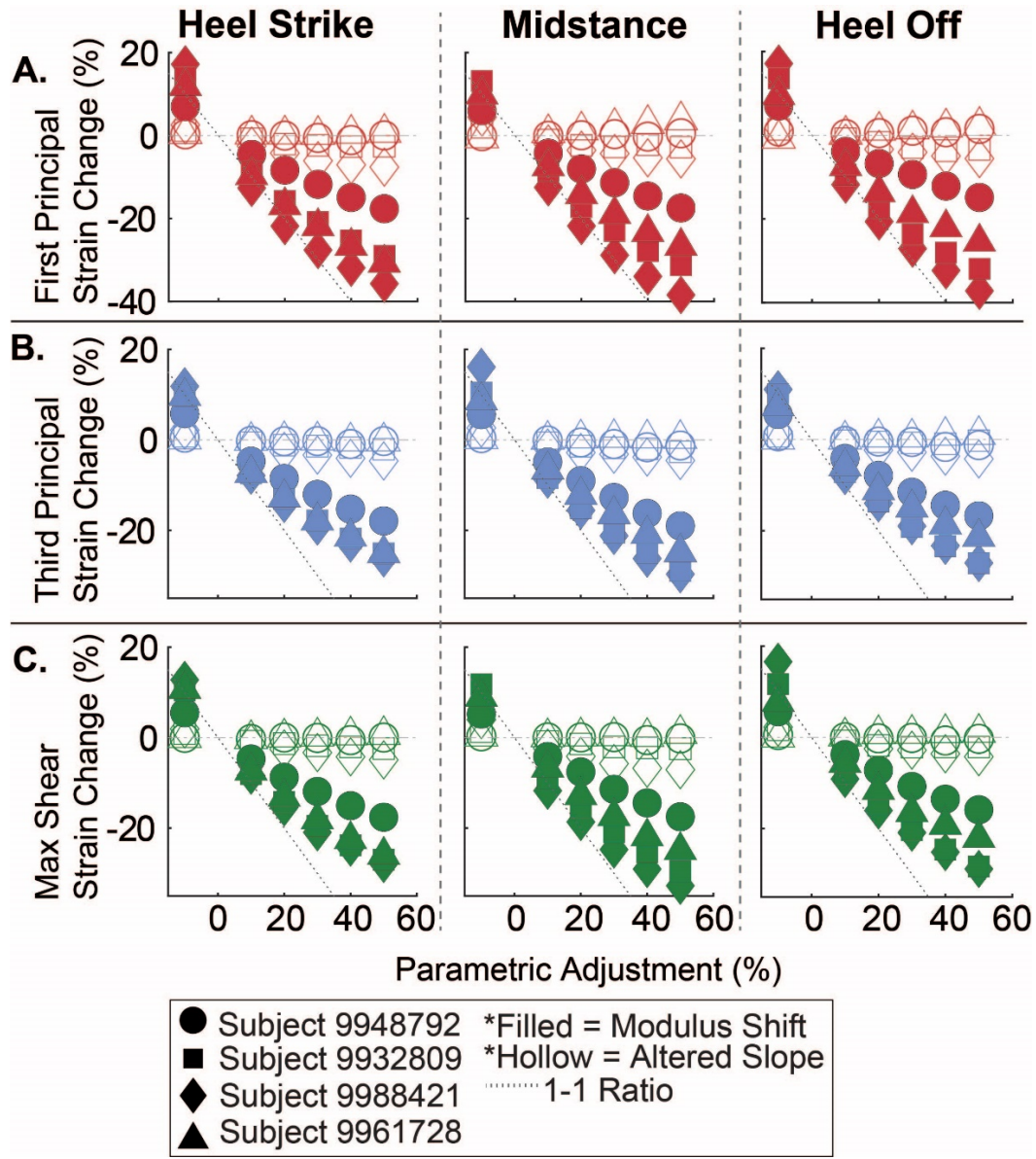

**Figure 6. Percent change in the key strain outputs from $T_2$-$E_D$ manipulation.** The percent change in the top 1% strain metrics at heel strike, midstance, and heel off was computed between the baseline simulation and simulations implementing 10% incremental changes to the $T_2$-$E_D$ relationship parameters. Solid shapes represent the modulus shift whereas hollow shapes as a result from the changes to the slope relationship.

## DISCUSSION

In this study, we first analyzed the texture of $T_2$ values assigned to the cartilage as a result from two voxel-to-element mapping approaches, nearest-neighbor approach and volume-weighted, comparing the roughness and RMS of assigned $T_2$ values. Selection of the volume-weighted approach then enabled our primary objective of evaluating the sensitivity of the $T_2$-informed FE models to changes to the linear $T_2$-$E_D$ relationship. We then compared changes to key stress and strain outputs from $T_2$-informed FE models after perturbing either the intercept or slope of this relationship.

We compared a volume-weighted voxel-to-element mapping that accounts for multiple partial overlapping voxels to the previously implemented nearest-neighbor approach (Lampen*, et al.* 2023). The volume-weighted approach produced a lower roughness and RMS of assigned $T_2$ compared to the nearest-neighbor approach. The nearest-neighbor approach was likely more sensitive to noisy voxels, potentially leading to higher $T_2$ differences between adjacent elements. On the other hand, the volume-weighted approach effectively enabled averaging of the $T_2$ of the tissue in the overlapping volume. In addition, the nearest-neighbor approach implemented on with one subject led to a negative Jacobian convergence error, suggesting that the noisy $T_2$ mapping resulting from this approach produces models that may be more sensitive to convergence errors. As a result, although the nearest-neighbor approach was vastly faster, the sensitivity study was performed with the volume-weighted approach.

The primary objective of this study was to evaluate the sensitivity of key outputs of a $T_2$ informed FE model to changes in the $T_2$-$E_D$ relationship that defined articular cartilage material properties. Shifting the $E_D$ range, while maintaining the slope, led to large changes in the top stress and strain percentiles, but altering the slope between $E_D$ and $T_2$ had a negligible impact on the computed outputs (Figure 5 and 6; Supplemental Table 1 and 2). These results show that a -10% to 50% perturbation of the $T_2$-$E_D$ slope did not substantially change the computed responses. On the other hand, uncertainty in $E_D$ range can lead to an under- or over-estimation of the number of elements under high stress and strain and their magnitudes, reflecting the fundamental relationships between material properties and the stress-strain relationship observed in other sensitivity studies.

Ayobami, *et al.*, employed a linearly elastic model at near incompressibility and simulated tibial compressive loading in a mouse finite element model (Ayobami, *et al.* 2022). A 100% increase in Young's Modulus resulted in a 13% increase in maximum compressive principal stress and a 41.4% decrease in maximum compressive principal strain. While our results show a lower sensitivity compared to Ayobami *et el.*, (Ayobami, *et al.* 2022), this is likely due to the range that $E_D$ changes. With the linear $T_2$-$E_D$ relationship, a 10% modulus shift from the baseline $E_D$ range meant that every element's $E_D$ increased by 0.7 MPa. Therefore, elements with an initial 1 MPa modulus would increase by 70% whereas elements with an initial 8 MPa would increase only by 8.75%. Our results show an average 4.9% increase and an average 7.2% decrease in the top 1% of the third principal stress and strain, respectively between all four subjects at midstance (Supplemental Tables S2 and S3).

Unexpectedly, the model was not sensitive to -10% to 50% changes in the slope of the $T_2$-$E_D$ relationship (Figures 5 and 6). This is likely because as the $T_2$ - $E_D$ relationship rotates about the midpoint at (45 ms, 4.5 MPa), $T_2$ values less than 45 ms were assigned lower modulus, while $T_2$ greater than 45 ms were assigned a higher modulus. Therefore, the lack of change in mean $E_D$ resulting from the altered slope might explain the low sensitivity. Because noninvasive estimation of cartilage modulus is limited to research environments (Butz, *et al.* 2011), defining subject-specific $T_2$-$E_D$ relationships is clinically infeasible. However, our results suggest that defining subject-specific relationships may not be necessary, since stress and strain outputs are not highly sensitive to the $T_2$-$E_D$ slope.

The sensitivity to changes in modulus range and slope differed between subjects (Figures 5 and 6). Because subjects were not scanned during load bearing, manual adjustments to femoral and tibial alignment were necessary. These model-specific joint alignment adjustments, combined with subject-specific anatomies, potentially explain the sensitivity differences among subjects. Zheng *et al.*

(Zheng, *et al.* 2017) reported up to a 70% increase in the lateral femoral cartilage maximum compressive stress as a result from a 6.6° valgus positioning compared to a baseline 1.3° varus positioning. Work by Anderson *et al.* (Anderson, *et al.* 2010) and Li *et al.*, (Li, *et al.* 2001) highlight FE model sensitivity to differences in cartilage morphology. Peak pressures in models with a constant acetabular cartilage thickness are greater than those of subject-specific models (Anderson, *et al.* 2010). Increased surface pressure, Von Mises stress, and hydrostatic pressure also increase in models with 10% thinner cartilage (Li, *et al.* 2001). Although we observe subject-to-subject variance in sensitivity, these other studies suggests that geometry and alignment may drive these uncertainties more than the $T_2$-$E_D$ relationship.

In this study, we implemented an isotropic, compressible neo-Hookean model because of the limited availability of experimental data directly linking MRI biomarkers to cartilage mechanics. However, collagen fibers and proteoglycans dictate a nonlinear tensile and compressive mechanical behavior in articular cartilage (Broom and Poole 1983, Chahine, *et al.* 2004, Williamson, *et al.* 2003). In this study, $E_D$ and $v$ were used to represent the averaged continuum response of cartilage, but the neo-Hookean model does not separate the individual contributions of extracellular matrix components that are known to change in early OA (Saarakkala, *et al.* 2010); nor does this hyperelastic material include viscoelastic and poroelastic behaviors of cartilage. The links between diffusion MRI and collagen orientation (Filidoro, *et al.* 2005, Raya, *et al.* 2011) and multiple MR relaxometry metrics to tissue content (Akella, *et al.* 2001, Keenan, *et al.* 2011, Wheaton, *et al.* 2004) suggest that MRI-based correlates to material properties may better account for tissue heterogeneity and focal changes to cartilage associated with the early stages of OA. Despite the limitations of the constitutive model chosen, this sensitivity study demonstrates the feasibility and limits of applying $T_2$ informed elastic properties derived from cadaveric experiments to FE models of different subjects, supporting our general strategy of incorporating subject-specific and spatially heterogeneous MRI biomarkers as correlates to tissue material properties.

In conclusion, this work highlights the importance of using an appropriately tuned modulus range when defining finite element models using linear relationships between $E_D$ and $T_2$ especially if using stress or strain cutoffs are intended to predict tissue damage. The insensitivity to slope suggests that the $T_2$ to $E_D$ relationship derived from limited experimental studies may well be applicable to a broader range of subjects. Although efforts to noninvasively determine true, subject-specific material properties warrants continued research, our analysis of the impact of $T_2$ to $E_D$ slope and intercept uncertainty on FE model outcomes supports the use of linear relationships derived from experimental studies for subject-specific FE modeling.

## DATA AVAILABILITY

The codes for $T_2$ smoothing and calculation and simulation pre-processing are available on GitHub: https://github.itap.purdue.edu/Chan-Lab/T2-Informed-Modeling.git. Imaging data (Peterfy, *et al.* 2008) and FE meshes (Rodriguez-Vila, *et al.* 2017) are available through the cited original sources.

## DISCLOSURE OF INTEREST

The authors report there are no competing interests to declare.

## ACKNOWLEDGEMENTS

This work was funded in part by NSF Award 2149946. We thank Nathan Lampen for his helpful advice in initiating these studies.

# Supplemental Information

**RESULTS**

## Subject 9948792

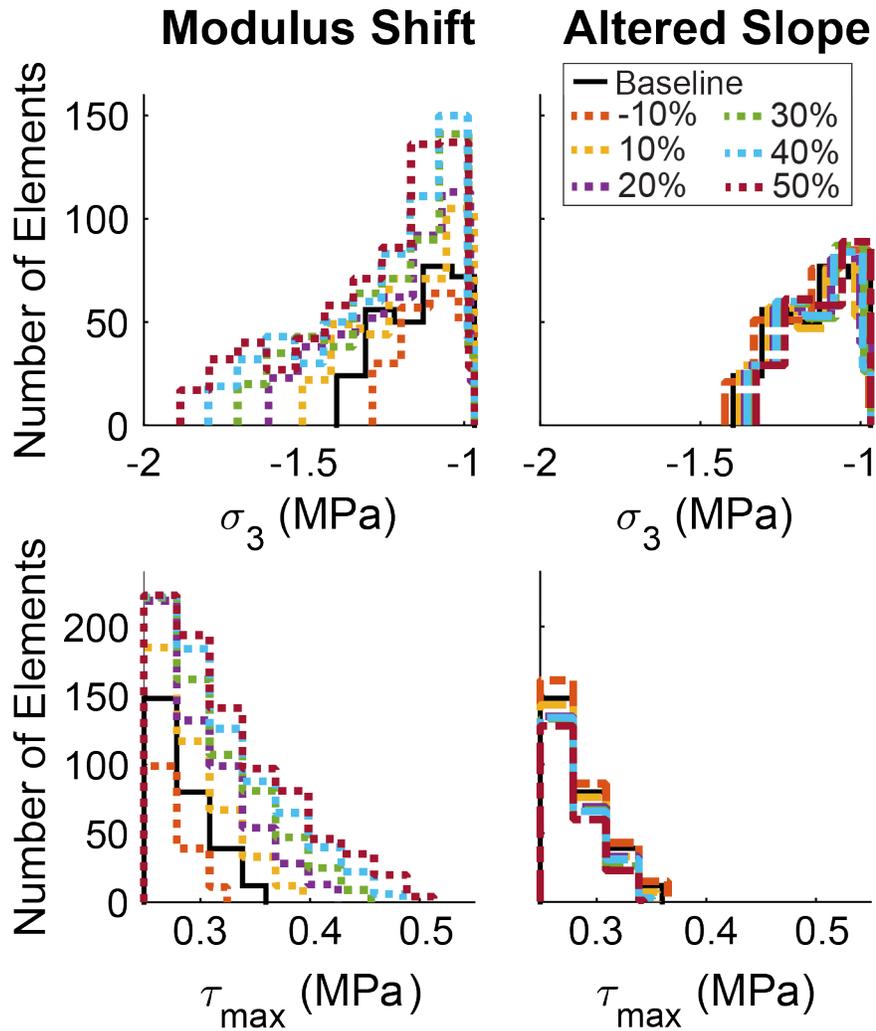

**Figure 1. Mechanical response as a result from the modulus shift (left column) and altered slope (right column).** The histogram details the number of elements above the baseline simulation top 1% value.

# Subject 9932809

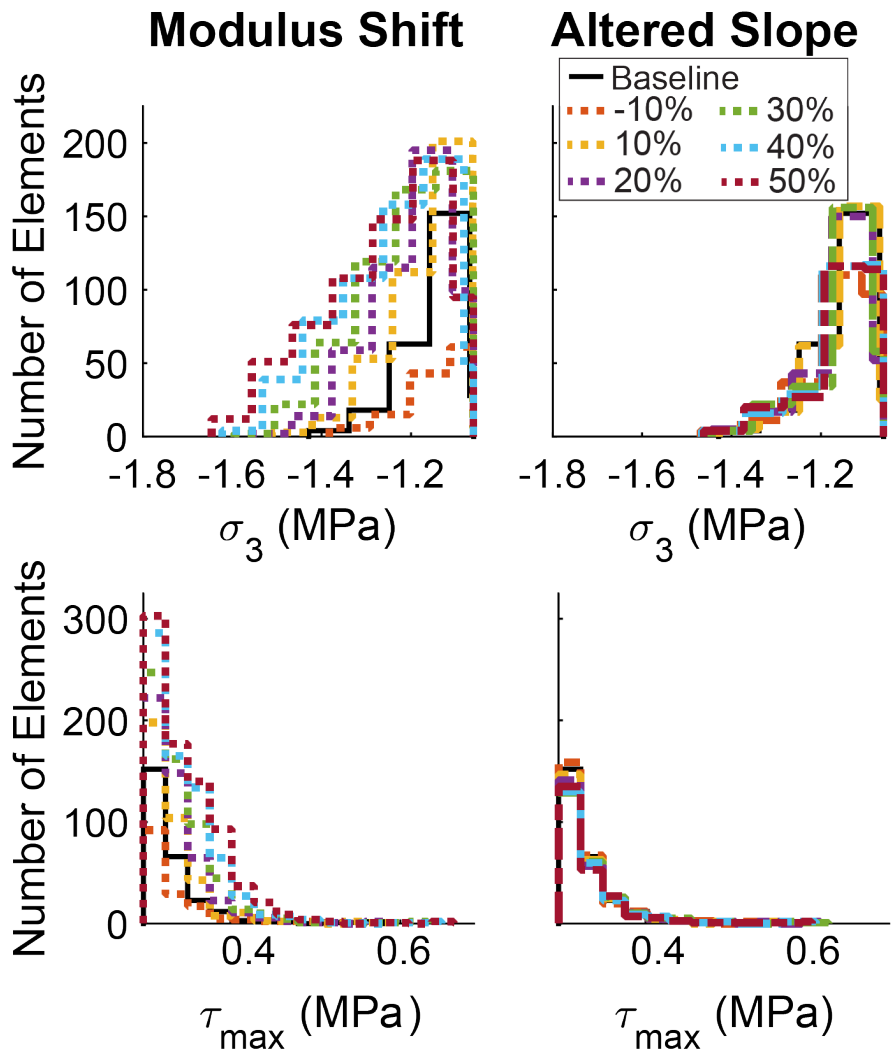

**Figure 2**: **Mechanical response as a result from the modulus shift (left column) and altered slope (right column).** The histogram details the number of elements above the baseline simulation top 1% value.

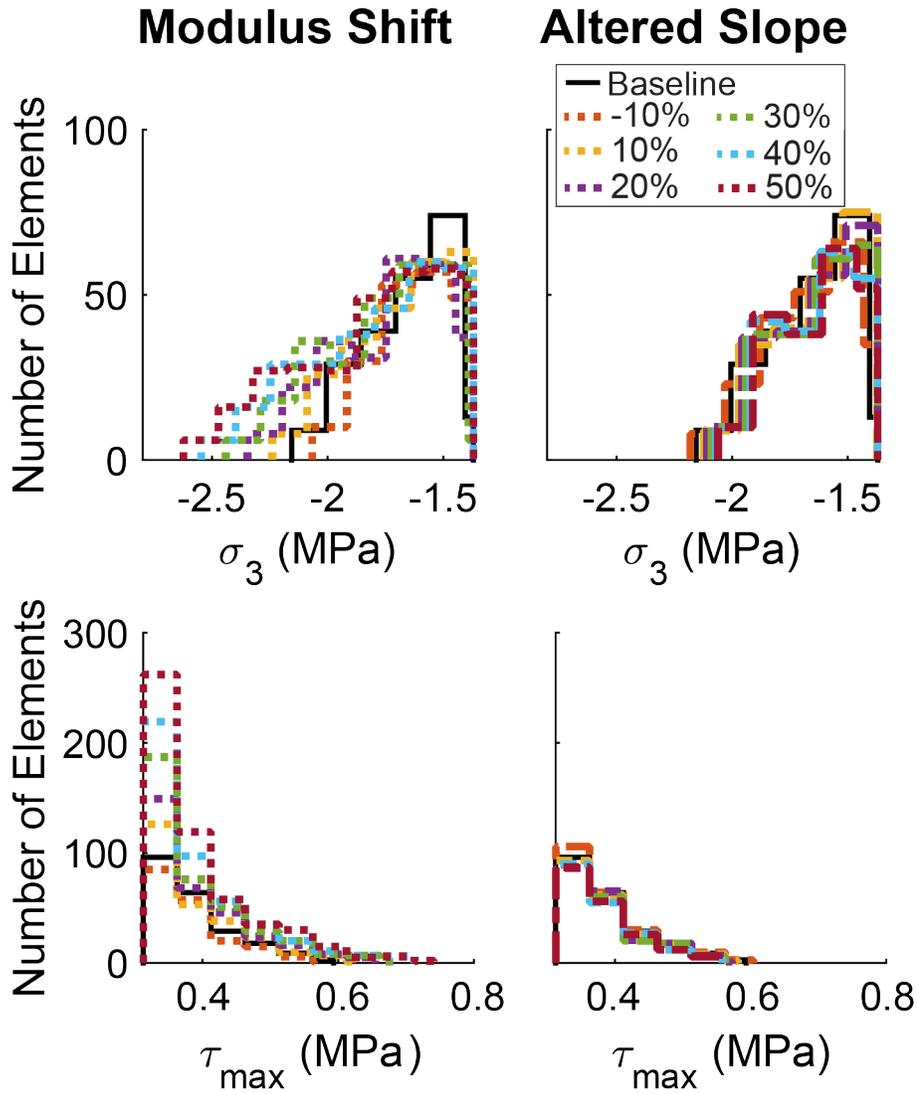

**Figure 3: Mechanical response as a result from the modulus shift (left column) and altered slope (right column).** The histogram details the number of elements above the baseline simulation top 1% value.

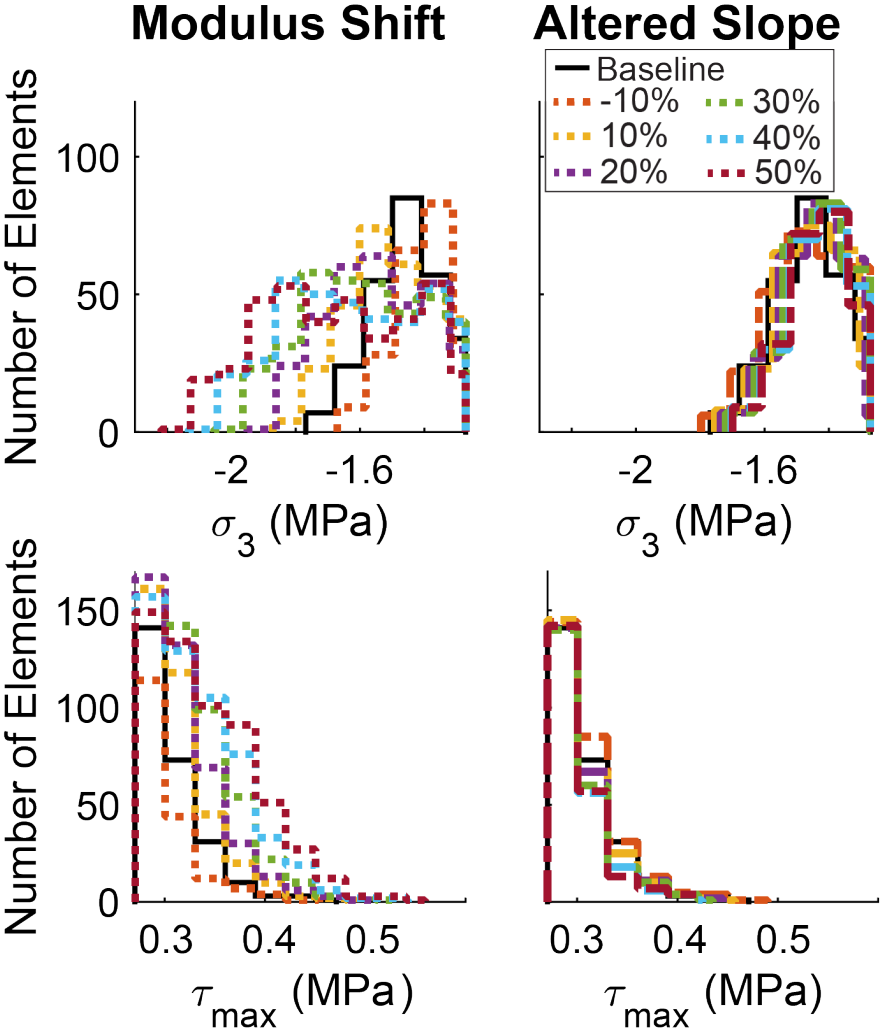

**Figure 4: Mechanical response as a result from the modulus shift (left column) and altered slope (right column).** The histogram details the number of elements above the baseline simulation top 1% value.

**Table 1: Mean percent change in stress outputs.** The mean and standard deviation from the top 1% stress outputs was calculated from all subjects as a result from the $T_2$-$E_D$ manipulations.

|  | Heel Strike | | Midstance | | Heel Lift | |
|---|---|---|---|---|---|---|
|  | Modulus Shift | Altered Slope | Modulus Shift | Altered Slope | Modulus Shift | Altered Slope |
| Mean First Principal Stress Change (%) | | | | | | |
| -10% | -5.4±0.9 | 0.0±1.4 | -6.9±1.9 | 0.4±0.3 | -7.3±0.9 | 0.4±0.3 |
| 10% | 5.3±1.4 | 0.2±0.2 | 6.8±1.4 | -0.1±0.5 | 6.8±1.4 | -0.2±0.8 |
| 20% | 10.7±0.9 | -0.1±0.7 | 12.5±2.76 | -0.0±0.8 | 13.7±1.0 | -0.3±1.4 |
| 30% | 15.9±3.1 | 0.2±0.7 | 18.6±5.1 | -0.1±1.1 | 20.6±3.1 | 0.2±1.8 |
| 40% | 20.0±3.9 | 0.3±0.4 | 24.1±6.2 | 0.4±0.8 | 26.0±4.0 | 0.3±1.6 |
| 50% | 25.4±4.8 | 0.1±0.7 | 29.9±8.6 | 0.1±1.4 | 32.1±4.8 | 0.00±1.9 |
| Mean Third Principal Stress Change (%) | | | | | | |
| -10% | -5.7±1.2 | 0.7±0.3 | -5.1±2.2 | 0.9±0.8 | -5.8±1.2 | 1.0±0.4 |
| 10% | 5.5±1.2 | -0.5±0.4 | 4.9±2.0 | -0.4±0.5 | 5.0±1.2 | -0.9±0.6 |
| 20% | 10.6±1.7 | -1.1±0.7 | 9.0±3.5 | -0.8±0.9 | 10.1±1.7 | -1.3±0.9 |
| 30% | 15.3±2.4 | -1.4±1.2 | 13.1±5.1 | -1.3±1.5 | 14.3±2.4 | -1.9±1.3 |
| 40% | 20.0±3.0 | -1.6±0.9 | 16.7±6.5 | -1.6±1.6 | 18.7±3.0 | -2.0±1.4 |
| 50% | 24.7±3.5 | -2.2±1.4 | 20.2±7.7 | -2.1±1.9 | 22.7±3.5 | -2.5±1.9 |
| Mean Max Shear Stress Change (%) | | | | | | |
| -10% | -6.9±1.7 | 1.0±0.4 | -6.2±1.7 | 1.1±0.4 | -7.2±1.6 | 1.5±0.4 |
| 10% | 6.1±1.7 | -1.1±0.2 | 6.4±1.6 | -0.8±0.5 | 6.9±1.7 | -1.1±0.6 |
| 20% | 12.4±2.7 | -1.9±0.5 | 11.8±3.4 | -1.5±0.7* | 13.2±2.7 | -1.8±0.7 |
| 30% | 18.3±3.5 | -2.3±1.1 | 17.5±4.5 | -2.2±1.0 | 19.2±3.5 | -2.5±1.2 |
| 40% | 24.1±4.5 | -2.6±0.9 | 22.5±5.4 | -2.7±0.7 | 25.2±4.5 | -2.7±1.0 |
| 50% | 29.8±5.4 | -3.3±1.5 | 27.5±6.7 | -3.5±1.4 | 30.8±5.4 | -3.5±1.3 |

**Table 2: Mean strain percent change.** The mean and standard deviation from the top 1% strain outputs was calculated from all subjects as a result from the $T_2$-$E_D$ manipulations.

|  | Heel Strike | | Midstance | | Heel Off | |
|---|---|---|---|---|---|---|
|  | Modulus Shift | Altered Slope | Modulus Shift | Altered Slope | Modulus Shift | Altered Slope |
| Mean First Principal Strain Change (%) | | | | | | |
| -10% | 12.5±4.3 | 1.1±1.0 | 13.0±7.3 | 1.4±2.3 | 11.8±4.5 | 0.8±1.1 |
| 10% | -8.9±3.4 | -0.7±1.3 | -8.2±3.4 | -0.6±1.2 | -8.1±34 | -0.6±1.0 |
| 20% | -15.7±5.6 | -1.5±2.1 | -15.3±5.8 | -1.1±2.2 | -14.6±6.0 | -0.9±2.1 |
| 30% | -20.5±6.6 | -2.2±2.8 | -20.5±7.3 | -1.2±2.9 | -19.7±7.7 | -1.1±2.8 |
| 40% | -24.7±7.1 | -2.6±3.2 | -24.9±8.2 | -1.3±3.4 | -23.7±8.7 | -1.5±3.3 |
| 50% | -28.3±7.6 | -2.4±3.8 | -28.5±8.8 | -1.1±4.0 | -27.4±9.6 | -1.3±3.8 |
| Mean Third Principal Strain Change (%) | | | | | | |
| -10% | 9.3±2.5 | 0.9±0.9 | 10.2±4.4 | 1.0±1.1 | 8.4±2.5 | 0.8±0.7 |
| 10% | -6.9±1.4 | -0.6±0.6 | -7.2±1.8 | -0.6±0.8 | -6.3±1.7 | -0.4±0.5 |
| 20% | -12.1±2.5 | -1.0±1.1 | -13.0±3.0 | -1.1±1.3 | -11.6±2.8 | -0.7±1.2 |
| 30% | -16.7±3.2 | -1.3±1.5 | -17.7±3.9 | -1.4±1.5 | -16.2±3.6 | -0.9±1.4 |
| 40% | -20.4±3.4 | -1.8±1.6 | -22.1±4.5 | -1.8±1.7 | -19.9±4.3 | -1.3±1.9 |
| 50% | -23.6±3.8 | -1.8±2.1 | -25.6±4.9 | -2.1±2.0 | -23.1±5.0 | -1.2±2.2 |
| Mean Max Shear Strain Change (%) | | | | | | |
| -10% | 9.9±3.1 | 0.7±0.9 | 11.6±6.5 | 1.5±1.8 | 10.4±4.8 | 1.1±1.0 |
| 10% | -7.1±1.7 | -0.7±0.5 | -7.9±3.2 | -0.8±1.1 | -6.6±2.4 | -0.5±0.8 |
| 20% | -12.9±2.9 | -0.9±1.2 | -13.6±5.7 | -1.3±2.0 | -12.4±3.9 | -0.9±1.5 |
| 30% | -17.6±4.0 | -1.2±1.7 | -18.6±5.7 | -1.8±2.8 | -17.1±4.7 | -1.1±1.7 |
| 40% | -21.4±4.3 | -1.6±2.0 | -22.7±6.2 | -2.3±3.1 | -20.6±5.4 | -1.4±2.0 |
| 50% | -24.6±4.8 | -1.7±2.5 | -26.2±6.6 | -2.2±3.5 | -23.8±6.1 | -1.4±2.3 |